\documentclass[class=preprint]{nature}
\usepackage{graphicx}
\graphicspath{{../../Figures/}}
\usepackage[usenames,dvipsnames]{color}
\usepackage[colorlinks=true,citecolor=blue,linkcolor=blue,urlcolor=blue]%
{hyperref}
\usepackage[a4paper]{geometry}
\usepackage{fixltx2e}
\usepackage{upgreek}
\usepackage{amsmath}
\usepackage{amssymb}
\usepackage{amsthm}
\usepackage{mathrsfs}
\usepackage{amsbsy}
\usepackage{amstext}
\usepackage{xcolor}
\usepackage{float}
\usepackage{multirow}
\usepackage{wasysym}
\usepackage{mathtools}
\usepackage{subfigure}
\usepackage{mhchem}
\usepackage{amsfonts}
\usepackage{dcolumn}
\usepackage{placeins}
\usepackage{float}
\usepackage{caption}
\usepackage[right]{lineno}
\usepackage{fancyhdr}
\usepackage{bbold,bm}
\usepackage{pdfpages}
\usepackage[utf8]{inputenc}
\usepackage[T1]{fontenc}%
\usepackage{setspace}
\providecommand{\U}[1]{\protect\rule{.1in}{.1in}}
\DeclareCaptionLabelSeparator{vline}{$\boldsymbol{:}$}
\title{Terahertz field-driven magnon upconversion in an antiferromagnet}
\author{Zhuquan Zhang$^{1\dagger}$, Frank Y. Gao$^{2\dagger}$, Yu-Che Chien$^1$, Zi-Jie Liu$^1$, Jonathan B. Curtis$^3$, Eric R. Sung$^1$, Xiaoxuan Ma$^4$, Wei Ren$^4$, Shixun Cao$^{4*}$, Prineha Narang$^3$, Alexander von Hoegen$^5$, Edoardo Baldini$^{2*}$, and Keith A. Nelson$^{1*}$ \\
\normalsize{$^1$Department of Chemistry, Massachusetts Institute of Technology, Cambridge, Massachusetts, USA, 02139 }\\
\normalsize{$^2$Department of Physics, The University of Texas at Austin, Austin, Texas, USA, 78712 } \\
\normalsize{$^3$College of Letters and Science, University of California, Los Angeles, California, USA, 90095}\\
\normalsize{$^4$Department of Physics, Materials Genome Institute and International Center for Quantum and Molecular Structures, Shanghai University, Shanghai, 200444, China} \\ 
\normalsize{$^5$Department of Physics, Massachusetts Institute of Technology, Cambridge, Massachusetts, USA, 02139 }\\ 	
\normalsize{$^{*}$E-mail: kanelson@mit.edu, edoardo.baldini@austin.utexas.edu \& sxcao@shu.edu.cn} \\
\normalsize{$^\dagger$These authors contributed equally to this work} \\
\\
}
\begin{document}

\maketitle

\section*{Abstract}
Tailored light excitation and nonlinear control of lattice vibrations have emerged as powerful strategies to manipulate the properties of quantum materials out of equilibrium. Generalizing the use of coherent phonon-phonon interactions to nonlinear couplings among other types of collective modes would open unprecedented opportunities in the design of novel dynamic functionalities in solids. For example, the collective excitations of magnetic order -- magnons -- can carry information with little energy dissipation\cite{chumak2015magnon}, and their coherent and nonlinear control would provide an attractive route to achieve collective-mode-based information processing and storage in forthcoming spintronics and magnonics.\cite{hortensius2021coherent, bae2022exciton} Here, we discover that intense terahertz (THz) fields can initiate processes of magnon upconversion mediated by an intermediate magnetic resonance. By using a suite of advanced spectroscopic tools, including a newly demonstrated two-dimensional (2D) THz polarimetry technique enabled by single-shot detection, we unveil the unidirectional nature of coupling between distinct magnon modes of a canted antiferromagnet. Calculations of spin dynamics further suggest that this coupling is a universal feature of antiferromagnets with canted magnetic moments. These results demonstrate a route to inducing desirable energy transfer pathways and THz-induced coupling between coherent magnons in solids and pave the way for a new era in the development of ultrafast control of magnetism.
\newpage

\section*{Main Text}
 
Nonlinear dynamics govern many exotic phenomena in complex systems, ranging from chaos\cite{strogatz2018nonlinear} and turbulence\cite{foias2001navier} to neural activity\cite{friston2001book}. In condensed matter physics, a notable manifestation of nonlinear dynamics is \textit{nonlinear phononics}. Intense terahertz (THz) or infrared laser pulses, derived from tailored light sources, can drive the collective vibrational modes of a lattice (phonons) into a coherent nonlinear regime, activating specific phonon-phonon interactions that would be silent otherwise.\cite{forst_nonlinear_2011, kozina2019terahertz} These couplings have been found to play a key role in many remarkable phenomena, including transient enhancement of superconductivity\cite{mankowsky2014nonlinear}, nucleation of macroscopic polarization\cite{li2019terahertz,nova2019metastable}, and manipulation of magnetic states\cite{nova_effective_2017, disa2020polarizing, afanasiev2021ultrafast, stupakiewicz2021ultrafast, juraschek2021sum}. In principle, similar coherent couplings are expected among other types of collective excitations, including magnons, plasmons, phasons, and amplitudons,\cite{michael2020parametric, juraschek2020parametric, curtis2022cavity} but their observation has thus far been hindered by weak interactions with light or fast decoherence. Addressing this challenge could establish an unprecedented paradigm in the nonequilibrium control of quantum materials and open the door to next-generation information technologies based on coupled coherent excitations.

Here we unveil a readily accessible coupling mechanism by studying the magnon modes of a canted antiferromagnet. We observe magnon upconversion processes mediated by nonlinear magnetic interactions. Unlike previous observations of magnon multi-harmonics\cite{kiselev2003microwave, lu2017coherent, kurihara_spin_nodate}, spin dynamics in the non-perturbative regime\cite{mukai2016nonlinear, schlauderer2019temporal}, and nonlinear manipulation of spins by a THz electric field\cite{baierl2016nonlinear, mashkovich2019terahertz, schlauderer2019temporal}, our results involve coupling between distinct magnon modes.

To unravel the exotic physics of magnon upconversion for the first time, we performed a series of THz pumping experiments on the model canted antiferromagnet ErFeO$_3$ (EFO), which crystallizes in a distorted orthorhombic structure.  Below the N\'{e}el temperature $T_N=643$ K, the spins of neighboring Fe ions (i.e., $S_{1}$ and $S_{2}$) order antiferromagnetically, yet slightly cant due to the Dzyaloshinskii–Moriya interaction, yielding a net magnetic moment $\mathbf{M}$.  As a result, magnon excitation follows specific selection rules: the quasi-antiferromagnetic (qAFM) mode can only be resonantly driven by a THz field whose magnetic field component is oriented along $\mathbf{M}$ while the excitation of the quasi-ferromagnetic (qFM) mode requires a magnetic field {component} perpendicular to $\mathbf{M}$.\cite{yamaguchi2013terahertz,li2018observation} Under a weak perturbation, the qFM mode corresponds to a precession of the magnetization orientation, while the qAFM mode represents a periodic modulation of the magnetization amplitude (see Fig. 1a).

We begin by studying the driven magnon response at room temperature using THz electron spin resonance (ESR) polarimetry. We focus a linearly polarized, single-cycle THz pulse with a peak magnetic field $\mathbf{H}_{\textup{THz}}$ of $\sim$ 0.15 T onto a (010)-cut single-domain EFO crystal, which allows both magnon modes to be accessible. The magnetic field component of the THz pump pulse is first aligned with respect to either the $a$ ($\perp \mathbf{M}$) or $c$ ($\parallel\mathbf{M}$) crystallographic axis to selectively drive the qFM or the qAFM mode, respectively. To measure the THz-induced free-induction decay (FID) signals, $\mathbf{H}_{\textup{det}}$, of the driven modes, we use a recently developed single-shot detection method\cite{teo2015invited, gao2022high} which enables the rapid acquisition of THz waveforms with high signal-to-noise ratio. A wire-grid polarizer is placed in front of the detector to select the emitted THz field either parallel or perpendicular to that of the THz excitation pulses ($\mathbf{H}_{\textup{det}} \parallel$ $\mathbf{H}_{\textup{THz}}$ or $\mathbf{H}_{\textup{det}} \perp$ $\mathbf{H}_{\textup{THz}}$) (see Fig. 1b). In the linear response regime, driving a specific magnon mode along either the $a$ or $c$ crystallographic axis can generate FID signals only parallel to the pump magnetic field ($\mathbf{H}_{\textup{det}} \parallel$ $\mathbf{H}_{\textup{THz}}$) and none perpendicular to it ($\mathbf{H}_{\textup{det}} \perp$ $\mathbf{H}_{\textup{THz}}$). Figure 1c shows the time-domain FID signals for both excitation and detection configurations. As expected, each FID signal in the parallel configuration reveals the coherent excitation of a single magnon mode (see Figure 1c, top panel), and the Fourier spectra in Fig. 1d match the previously reported magnon frequencies at room temperature (qFM, 0.38 THz and qAFM, 0.67 THz)\cite{yamaguchi2013terahertz}. However, when the THz pump field drives the qFM mode ($\mathbf{H}_{\textup{THz}} \parallel a$), we clearly observe the excitation of the qAFM mode revealed by the emitted signal in the perpendicular polarization configuration, which is not predicted in linear response (see Figure 1c, dark red trace). Conversely, no qFM response is found when the THz magnetic field component is oriented along the $c$ axis to excite the qAFM mode (see Figure 1c, light grey trace).

We achieve a comprehensive understanding of the magnon symmetry by rotating the sample and repeating the same measurements every 5$^\circ$. The oscillation amplitudes of both magnon modes are extracted from the Fourier spectra of the FID signals for both detection configurations and plotted as a function of the azimuthal angle, $\theta$, i.e. the angle  between the THz magnetic field orientation and the $a$ axis in the $ac$ plane. For the qFM mode, the parallel polarimetry pattern shows two-fold rotation symmetry ($\cos^2\theta$) and reaches the largest amplitudes at $\theta$ = 0, with $\mathbf{H}_{\textup{THz}}$ and $\mathbf{H}_{\textup{det}}$ along the crystallographic $a$ axis. Since the perpendicular detection configuration eliminates any FID signal that arises when the THz magnetic field directly drives the qFM mode, the responses peak at 45$^\circ$ off-axes, proportional to $\left|\cos\theta \sin\theta\right|$. For the qFM mode, both polarimetry patterns are well captured by only considering the linear Zeeman interaction with the THz magnetic field.\cite{li2018observation, grishunin2021excitation} For the qAFM mode, the parallel polarimetry pattern also shows two-fold rotation symmetry ($\sin^2\theta$), with maximum signal amplitude when $\mathbf{H}_{\textup{THz}}$ and $\mathbf{H}_{\textup{det}}$ are along the $c$ axis. However, the perpendicular polarimetry pattern cannot be captured by the linear response alone. This is already apparent from the raw data in Fig. 1e: the qAFM response vanishes at 90$^\circ$ when the qAFM mode is directly driven by $\mathbf{H}_{\textup{THz}} \parallel c$ and its corresponding FID emission is blocked by the wire-grid polarizer, but the qAFM response is nonzero at 0$^\circ$ when $\mathbf{H}_{\textup{THz}} \parallel a$ drives the qFM mode. To express the full qAFM perpendicular detection pattern analytically, we perform a fit to the sum of both the linear response ($\left|\cos\theta \sin\theta\right|$) and the second-order magnon upconversion process driven by the excitation of the qFM mode ($\left|\cos^3\theta\right|$) and find an excellent match with the data. 

To further distinguish the nature of the resonant excitation and the nonlinear magnon upconversion process, we perform a 2D THz coherent spectroscopy measurement, which isolates nonlinear light-mode interactions with exceptional spectral resolution and has proven to be an effective tool to uncover the origins of various nonlinear responses.\cite{lu2017coherent,johnson2019distinguishing,mahmood2021observation, mashkovich2021terahertz, zhang2021nonlinear} By adding a second, time-delayed THz pulse (see Fig. 2a) and recording the time-domain THz field emission with variable inter-pulse delay, $\tau$, we extract the nonlinear THz signal generated by both pulses and perform a 2D Fourier transform to yield a frequency-frequency correlation map of the nonlinear magnonic response. The resulting 2D THz spectrum with $\mathbf{H}_{\textup{THz}} \parallel a$  detected in the perpendicular configuration shows a strong cross-peak with excitation and emission frequencies corresponding to the qFM and qAFM magnon modes, respectively (see Fig. 2b). The peak amplitude also scales as the square of the pump magnetic field, as shown in Fig. 2c, which unambiguously confirms the nonlinear upconversion process: the emission of the qAFM magnon mode is only allowed once the qFM magnon mode is excited.

We then analyze this response through a \textit{2D THz polarimetry} measurement of this off-diagonal magnon cross-peak. Similar to our linear polarimetry measurements, we utilize our ability to rapidly measure time-dependent THz fields via our single-shot THz measurement technique to enable 2D THz polarimetry, in which we rotate the azimuthal angle $\theta$ (see Fig. 2a) and collect 2D spectra at 5$^\circ$ increments. The resulting polarimetry patterns of the magnon cross-peak amplitude at ($\Omega_{qFM}$, $\Omega_{qAFM}$) are shown in Figs. 2c and 2d for both  parallel and perpendicular detection geometries.  The nonlinear upconversion signal shows two-fold symmetry in the perpendicular configuration and is strongest when $\mathbf{H}_{\textup{THz}} \parallel a$ driving the qFM mode, whereas in the parallel configuration it is a distorted clover pattern with maximum amplitude close to $30^\circ$. These polarimetry patterns can be well fit with functions proportional to $\left|\cos^2{\theta}\sin{\theta}\right|$ and $\left|\cos^3{\theta}\right|$, respectively. This indicates that the upconverted THz field emission at $\Omega_{qAFM}$ is second order, i.e. proportional to the square of the THz magnetic field component along the $a$ axis, while the emission orientation is that of the qAFM magnon, along the $c$ axis. We note that we do not observe the corresponding downconversion peak, i.e. qAFM $\rightarrow$ qFM, in the 2D THz spectrum by driving the qAFM mode alone, which further confirms this coupling to be asymmetric.  

Next, we study the robustness of the magnon upconversion process across a wide range of temperatures. As shown in Fig. 3a, EFO has a net magnetic moment along the crystallographic $c$ axis at room temperature. As the crystal is cooled, the material undergoes a spin reorientation transition (SRT) between 96-87 K, which continuously rotates the magnetic moment across the $ac$ plane, such that below 87 K the magnetic moment \textbf{M} aligns along the $a$ axis.\cite{gorodetsky1970sound,yamaguchi2013terahertz} This spin reorientation leads to a modification of the selection rules for the two magnon modes when $\textbf{k}_{\textup{THz}} \parallel b$, with $\mathbf{H}_{\textup{THz}} \parallel c$ driving the qFM mode and $\mathbf{H}_{\textup{THz}} \parallel a$ driving the qAFM mode. Accordingly, we align the THz magnetic field along the $a$ axis above the SRT and along the $c$ axis below the SRT to initiate the magnon upconversion process. We then detect the resulting qAFM coherence response using 2D THz spectroscopy in the perpendicular polarization configuration. The experimentally determined excitation and emission frequencies are shown in Fig. 3b with the corresponding 2D THz spectra at selected temperatures shown in Fig. 3c (additional spectra can be found in Supplementary Fig. S4). These experimental values are shown together with calculations of the qFM and qAFM mode frequencies (solid lines in Fig. 3b) and the 2D THz spectra (see Fig. 3c bottom row) derived by solving the Landau-Lifsitz-Gilbert (LLG) equation based on a single-site two-sublattice spin Hamiltonian with time-dependent effective magnetic fields (see Methods for simulation details). The excitation and emission frequencies of the magnon upconversion peak are well-matched with the calculated qFM and qAFM mode frequencies and those reported in the literature\cite{yamaguchi2013terahertz}. The qAFM frequency is seen to be relatively constant above the SRT but hardens below the transition while the qFM frequency softens near the SRT due to the cancellation of the second-order anisotropies along $a$ and $c$. Note that even in the region surrounding the SRT where $\Omega_{qAFM}/\Omega_{qFM} > 4$, the magnon upconversion process remains second-order, standing in stark contrast to other higher-order harmonic generation processes. 

From the current dataset, we can rule out the possibility that the qAFM mode is driven nonlinearly by the THz field itself and establish that the upconversion process is assisted by the intermediate magnon state. However, it remains unclear whether the coupling is activated by the combined motion of the mode and the THz field or driven by the magnon mode alone. Both mechanisms are described by the equations of motion up to second order with respect to the THz fields:
\begin{align}
	\dot{\gamma}=C_1 \gamma + C_2  \sigma^2 + C_3\sigma h,
\end{align}
where $C_1$, $C_2$, and $C_3$ are phenomenological coefficients; $\gamma$ and $\sigma$ are the qAFM and qFM coordinates; and $h$ is the magnetic field component of the THz pulse. Here, the first term represents the linear response of the qAFM mode directly driven by the light field. In addition to the linear term, there are second-order contributions that scale quadratically with the THz field. The second term (i.e., $\sigma^2$) only involves the driven qFM mode, which is analogous to the case of nonlinear phononics\cite{forst_nonlinear_2011}. The last term (i.e., $\sigma h$) is proportional to both the amplitude of the qFM magnon mode and the THz magnetic field, indicating that the coherence of the qAFM mode is induced by the joint effects of the intermediate magnon state and the THz field component at the difference frequency (i.e., $\Omega_{qAFM} - \Omega_{qFM}$). This is also analogous to the nonlinear phonon excitation process known as the infrared resonant Raman effect\cite{khalsa2021ultrafast}. Since both nonlinear effects originate from the initial excitation of the qFM magnon mode, the frequency correlation revealed by the off-diagonal 2D THz spectral peak cannot be directly distinguished between these two scenarios.

To address this question, we track the evolution of the magnon upconversion signal at different temperatures above the SRT. Figure 4a shows a finer temperature dependence of the upconversion amplitudes, along with the ratio of the qAFM magnon frequency to that of the qFM magnon mode (i.e., $\Omega_{qAFM}/\Omega_{qFM}$). The temperature dependence of the upconversion amplitudes exhibits distinct non-monotonic behavior, with nonzero signals at all temperatures and a maximum at around 210 K, where the phase-matching condition $2\Omega_{qFM} = \Omega_{qAFM}$ holds.  Indeed, this trend does not follow the simulated results considering either nonlinear excitation mechanism alone, but can be well reproduced by fitting to a combination of the two contributions (see Fig. 4b). This indicates that both mechanisms contribute to the overall nonlinear upconversion but diverge in that the field-driven process occurs with comparable amplitude across a wide range of temperatures whereas the magnonic sum-frequency excitation process is strongly enhanced close to 210 K where the phase-matching condition is met, and contributes non-negligibly but only at about one-fourth the amplitude of the field-driven process at temperatures well above 210 K and negligibly at temperatures well below 210 K. A comparable trend is also discernible below the SRT, as the upconversion signal reaches its peak at 25 K when the resonance condition $2\Omega_{qFM} = \Omega_{qAFM}$ is fulfilled (see Supplementary Fig. S8).

Our discovery reveals a hitherto-unknown pathway to drive magnon modes through their nonlinear interactions with the magnetic field of the THz pulses, and therefore expands the toolbox for controlling the properties of magnetic quantum materials out of equilibrium. For example, by using a multi-cycle THz field with the sum-frequency resonant condition satisfied (i.e., $2\Omega_{qFM} = \Omega_{qAFM}$), one can further strengthen the upconversion process. We also anticipate that this concept of inducing coherence transfer between distinct magnon modes will be generally applicable to many other magnetically ordered systems, including multiferroics, atomically thin antiferromagnets and coupled FM/AFM heterostructures, and will further extend the frontiers of spintronics and magnonics into the ultrafast nonlinear regime.

\newpage

\section*{Methods}
\noindent \textit{Sample Preparation} \\
Polycrystalline ErFeO$_3$ was synthesized by conventional solid state reaction using Er$_2$O$_3$ (99.9\%) and Fe$_2$O$_3$ (99.9\%) powders. According to their stoichiometric ratios, original reagents were weighed and then ground in an agate mortar with anhydrous ethanol. After grinding, the mixture was pre-sintered before being transferred back to the agate mortar and reground into powder. The powder was then pressed into round flakes using a mold and electric hydraulic press (YLJ-40T) before the second sintering and the regrinding. The polycrystalline powders were pressed into a bar by a Hydrostatic Press System (HP-M-SD-200) under an apparent pressure of 60 MPa, and then sintered again. Each sintering was performed for 1000 min at a temperature 1280 °C. 

\noindent Single-crystal ErFeO$_3$ samples were prepared in an optical floating zone furnace (FZ-T-10000-H-VI-P-SH, Crystal Systems Corp). The sintered polycrystalline rods were used both as the seed crystal and the feed rod. Single crystal growth was performed at a growth rate of 3 mm/h under an airflow of 2 L/min. A high-quality single crystal with a length of about 50 mm and a diameter of about 5 mm was obtained after about 17 hours of growth. The single crystal was then cut to obtain the (010)-cut sample with a thickness of $\sim$2 mm and lateral dimensions of $\sim$5 mm for THz measurements. Before each measurement, the sample was magnetized to reinforce the residual magnetization and ensure that a single magnetic domain is formed. The crystallographic orientation of the single crystal was determined with Laue diffraction (See Supplementary Fig. S2). 

\noindent \textit{Single-shot time-domain THz spectroscopy}\\
The majority of the output of a 1 kHz Ti:Sapphire laser amplifier (12 mJ, 35 fs) was split into two equal pulses that were variably delayed and overlapped with a tilted pulse-front geometry in a MgO:LiNbO$_3$ crystal\cite{yeh2008generation} to generate a pair of time-delayed THz pulses. These pulses were then focused onto the sample before being refocused onto a 2-mm ZnTe crystal to measure the THz waveform via electro-optic (EO) sampling. The remainder of the laser output was first expanded 9$\times$ and then reflected off an echelon mirror to generate a spatial array of 500 time-delayed and spatially shifted pulses. The pulses were focused along with the THz pulse(s) onto the EO crystal and then imaged onto a high-speed camera. The THz field-induced birefringence was detected by separating the probe array into two orthogonal polarizations with a balanced detection scheme. With this method, we can retrieve the time-domain THz signal covering 20 ps with a single THz pump laser shot while still operating at the full 1 kHz repetition rate of the laser amplifier. See Supplementary Fig. S1 for the experimental setup.

\noindent For THz FID measurements, one of the THz pump arms is chopped at 500 Hz (the other is blocked) and the corresponding THz FID signal averaged over 1000 shots (1 s) is recorded. For 2D THz measurements, a differential chopping scheme is used where the two THz pulses, $A$ and $B$, are chopped at 500 Hz and 250 Hz, respectively. The THz signals $\mathbf{H}_{AB}$, $\mathbf{H}_A$, $\mathbf{H}_B$, and $\mathbf{H}_0$, corresponding to traces taken with the respective pulse(s) unblocked, are averaged over 5000 shots (5 s) and recorded as the inter-pump pulse delay, $\tau$, is scanned between 3 and 20 ps. The time-domain nonlinear signal, $\mathbf{H}_{NL}(t,\tau)$, is then obtained via
\begin{equation}
	\mathbf{H}_{NL}(\tau,t) = \mathbf{H}_{AB}(\tau,t)  - \mathbf{H}_A(\tau,t)  - \mathbf{H}_B(t)  + \mathbf{H}_0(t),
\end{equation}
which is Fourier transformed with respect to both $t$ and $\tau$ to yield the 2D THz spectrum. 

\noindent {Single-shot detection has been utilized in cases where only a limited number of shots is required, such as in the study of photoinduced metastable phase transitions\cite{gao2022snapshots} and samples under extreme pulsed magnetic fields\cite{noe2016single}. In our measurements, single-shot detection of the signal field permits the 2D spectrum to be collected with only one time variable (the time interval between THz pulses) swept. The single-shot THz measurement method was developed earlier to facilitate 2D THz spectroscopy\cite{teo2015invited} and the approach has been applied recently\cite{duchi20212d, gao2022high}, but its full capabilities can be exploited most effectively as demonstrated here with the use of several hundred probe pulses to record the time-dependent field at the corresponding number of time points on each shot.

\newpage
\noindent \textit{THz ESR polarimetry}\\
\noindent For both 1D and 2D THz polarimetry measurements, the sample is placed on a motorized rotation stage and the corresponding THz signal is recorded as the azimuthal angle, $\theta$,  of the sample orientation relative to the incident THz polarization is rotated. A pair of wire grid polarizers, one placed before and one after the sample, is used to select the component of the transmitted/emitted THz field either parallel or perpendicular to the incident THz polarization. The data acquisition time for the complete set of 2D polarimetry measurements (144 2D spectra in all, collected with parallel and perpendicular detection configurations at 5$^\circ$ increments of the crystal orientation), was about 22 hours. We note that our previously reported 2D ESR spectra of magnons\cite{lu2017coherent}, collected without single-shot detection, required several days of data acquisition per spectrum; the present study would have required more than 2 years of data acquisition.

\noindent \textit{LLG simulations}\\
Numerical calculations of the THz-induced magnon dynamics are performed by solving the LLG equations based on the following Hamiltonian\cite{kampfrath2011coherent, lu2017coherent}
\begin{equation}
\begin{split}
		\mathcal{H} &= \mathcal{H}_0+\mathcal{H}_{Zeeman}\\
	 	&=nJ\mathbf{S}_{1}\cdot \mathbf{S}_{2}+n\mathbf{D}\cdot (\mathbf{S}_{1}\times \mathbf{S}_{2})-			\sum_{i=1,2}(K_{a}S_{ia}^2+K_{c}S_{ic}^2)\\
	 	&\hspace{12pt}-\gamma[\mathbf{H}_A(\tau,t)+\mathbf{H}_B(t)]\cdot(\mathbf{S}_{1}+\mathbf{S}_{2}).
\end{split}
\end{equation}
where $J$ is the Heisenberg exchange constant, $\mathbf{D}$ is the antisymmetric exchange constant, $K_a$ and $K_c$ are components of the onsite anisotropy along the $a$ and $c$ axis, respectively, and $n=6$ is the number of neighboring spins. The additional Zeeman term describes the interaction between the THz pulse's magnetic field and sublattice spins. The magnetic fields used here ($\mathbf{H}_A$ and $\mathbf{H}_B$) are the second derivatives of a Gaussian function, which ensure that the temporal integration of the pulses reduces to zero.  $\gamma=g\mu_B/ \hbar$ is the gyromagnetic ratio and $g=2$ is the g-factor. An equation of motion can be derived for each sublattice spin $\mathbf{S}_i$ $(i=1,2)$
\begin{equation} 
	\frac{d\mathbf{S}_i}{dt}=\frac{\gamma}{1+\alpha^2}[\mathbf{S}_i \times \mathbf{H}_i^{eff}+\frac{\alpha}{|\mathbf{S}_i|}\mathbf{S}_i \times (\mathbf{S}_i \times \mathbf{H}_i^{eff})],
\end{equation}
where $\alpha$ is a phenomenological Gilbert damping constant that accounts for energy dissipation, $\mathbf{H}_i^{eff}$ is the effective magnetic field for each lattice $i$, which can be calculated as $\mathbf{H}_i^{eff}=-\frac{1}{\gamma}\frac{\partial \mathcal{H}}{\partial \mathbf{S}_i}$. 

To account for the temperature dependence of the qFM and qAFM modes, we use parameters defined in Table S1. To simulate the 2D spectra, we solve for the time evolution of each sublattice spin $\mathbf{S}_i$ as a function of both $\tau$ and $t$ at each temperature and then extract the nonlinear response of the magnetization $\mathbf{M}=\mathbf{S}_1+\mathbf{S}_2$ generated by both magnetic fields. Performing 2D Fourier transforms yields the simulated 2D THz spectra (See Supplementary Fig. S5). The detection configurations are chosen to be the same as those used in the experiments. The simulated polarimetry patterns are shown in Supplementary Fig. S6.

\noindent\textbf{Data availability}
All data that support the findings of this study are available from the corresponding authors on reasonable request.

\noindent\textbf{Code availability}
All LLG simulation codes are available from the corresponding authors on reasonable request.

\noindent\textbf{Acknowledgments} 
Z.Z., Z.-J.L., E.R.S. and K.A.N acknowledge support from the U.S. Department of Energy, Office of Basic Energy Sciences, under Award No. DE-SC0019126. F.Y.G. and E.B. acknowledge support from the Robert A. Welch Foundation (grant F-2092-20220331). Y.-C.C. acknowledges direct funding from the MIT UROP. S.C. and W.R. acknowledge support from the Science and Technology Commission of Shanghai Municipality (No.21JC1402600) and the National Natural Science Foundation of China (NSFC, Nos.12074242, 12074241). Work by JC and PN was partially supported by the Department of Energy, Photonics at Thermodynamic Limits Energy Frontier Research Center, under Grant No. DE-SC0019140 and by the Quantum Science Center (QSC), a National Quantum Information Science Research Center of the U.S. Department of Energy (DOE). P.N. acknowledges support as a Moore Inventor Fellow through Grant No. GBMF8048 and gratefully acknowledges support from the Gordon and Betty Moore Foundation.

\noindent\textbf{Author contributions} Z.Z. conceived the study; Z.Z. and F.Y.G. designed and performed the experiments and analyzed the data, supported by Z.-J.L. and Y.-C.C.; Z.Z., Y.-C.C, and F.Y.G. performed spin dynamics simulations, supported by J.B.C. and E.R.S.; X.M. grew and cut the high quality single crystals used in the experiments under the guidance of W.R. and S.C.; Z.Z., F.Y.G., J.B.C., P.N., A.v.H., E.B., and K.A.N. interpreted the data; Z.Z., F.Y.G., E.B., A.v.H. and K.A.N. lead the manuscript preparation with input from all the authors; K.A.N. and E.B. supervised the project.     

\noindent\textbf{Competing interests} The authors declare no competing interests.

\section*{References}

\footnotesize
\bibliographystyle{naturemag}
\bibliography{paper}
\newpage
\normalsize

\FloatBarrier

\begin{figure}
	\centering
	\includegraphics[width=0.95\linewidth]{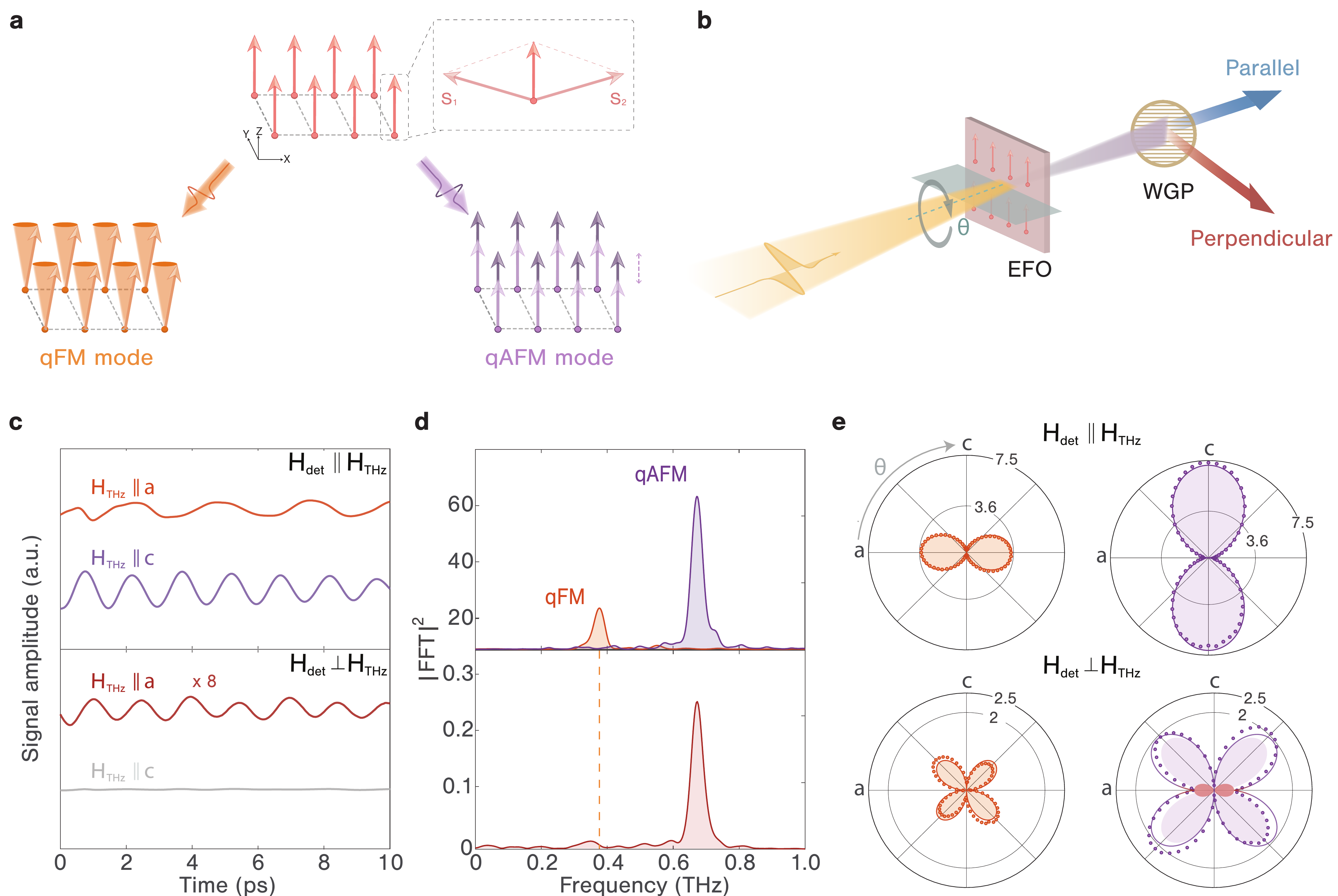} 
	\caption{\label{fig:Fig1}
	\textbf{THz field-driven magnon responses and ESR polarimetry on ErFeO$_3$ at room temperature. a,} A canted antiferromagnet with two sublattice spins possesses two magnetic-dipole allowed zone-center magnon modes: the qFM mode can be interpreted as a precession of the net magnetization and the qAFM mode can be understood as an amplitude oscillation of the magnetization. \textbf{b,} Schematic representation of the experimental setup for the time-domain FID and polarimetry measurements. A wire grid polarizer (WGP) is used to select for either the
parallel (blue) or perpendicular (red) polarized THz field emission. \textbf{c,} FID signals corresponding to excitation of either the $a$ axis qFM mode or the $c$ axis qAFM mode for both parallel and perpendicular signal detection configurations. \textbf{d,} Fourier transforms of the FID signals in \textbf{c}. The power spectra of the qFM (orange) and qAFM (purple) modes driven by direct THz magnetic excitations are shown at the top. The bottom power spectrum shows the qAFM mode response driven by the excitation of the qFM mode (red). \textbf{e,} (top) Parallel polarimetry patterns of the qFM (left) and qAFM (right) mode amplitudes. The experimental data (circles) of the qFM (qAFM) mode are fitted to functions of the form $\cos^2\theta$ ($\sin^2\theta$). (bottom) Perpendicular polarimetry patterns of the qFM (left) and qAFM (right) mode amplitudes. The responses are plotted on the same scale with their relative amplitudes labeled. The experimental data of the qFM mode are fitted to functions of the form $\cos\theta \sin\theta$, but the qAFM mode responses are fitted to a sum of the form $\left|\cos\theta \sin\theta\right|$ (light purple shaded area) and $\left|\cos^3\theta\right|$ (light red shaded area). The numerical values shown indicate relative amplitudes.}
\end{figure}

\begin{figure}
	\centering
	\includegraphics[width=0.95\linewidth]{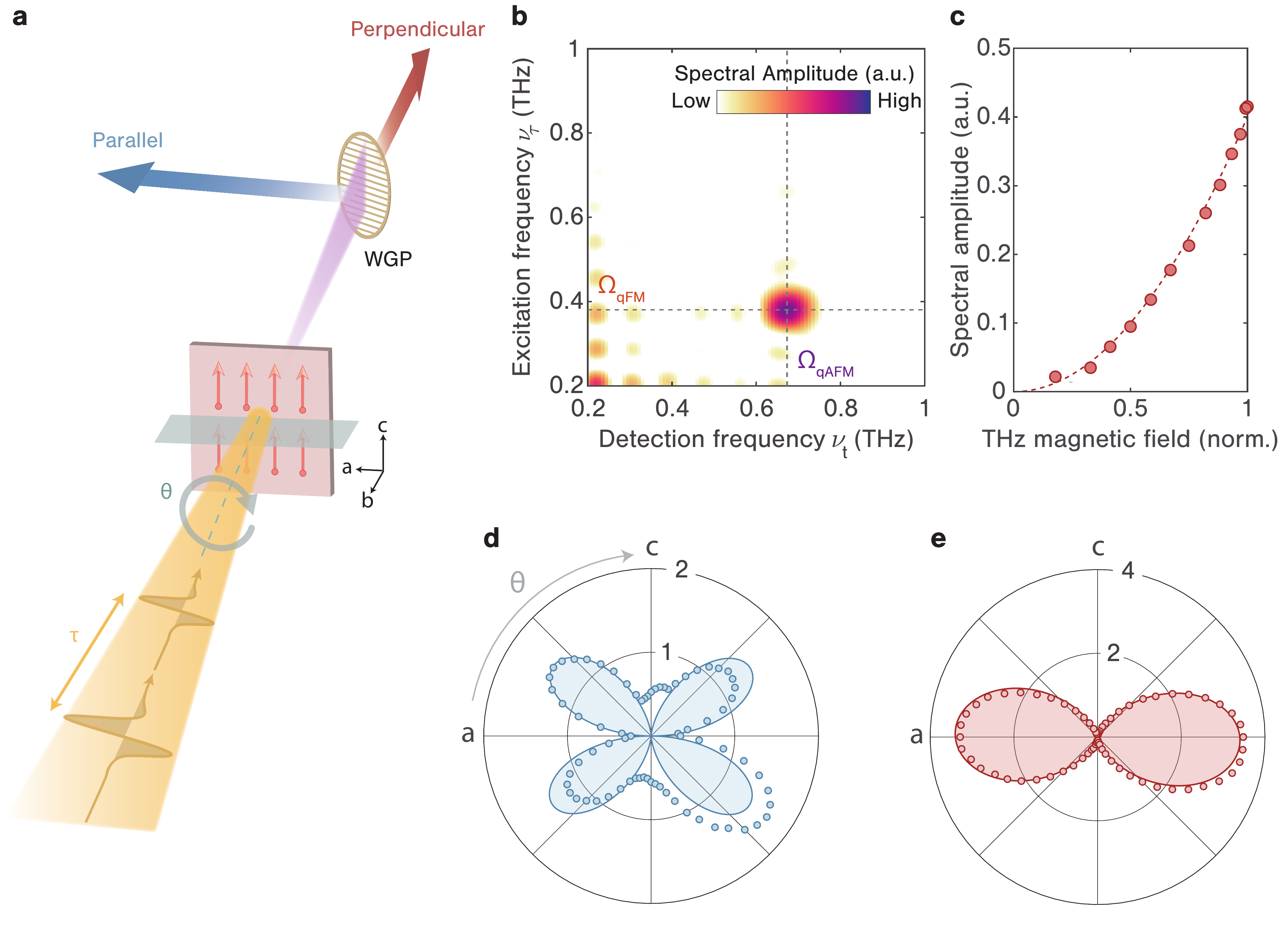}
	\caption{\label{fig:Fig2}  
	\textbf{2D THz ESR spectroscopy of magnon upconversion signal at room temperature. a,} Schematic illustration depicting the excitation scheme in the 2D THz ESR spectroscopy setup. \textbf{b,} Room temperature 2D THz spectra collected in the perpendicular detection geometry with $\mathbf{H}_{\textup{THz}} \parallel c$ showing the strong off-diagonal magnon upconversion peak. \textbf{c,} Dependence of the magnon upconversion signal amplitude on the pump magnetic field with H$_{\textup{det}} \perp $ H$_{\textup{THz}}$. The data are fitted with a quadratic function (dashed line). \textbf{d,} and \textbf{e,} Parallel (blue) and perpendicular (red) polarimetry patterns showing the amplitude of the 2D THz upconversion peak collected upon rotation of the azimuthal angle, $\theta$, between $\mathbf{H}_{\textup{THz}}$ and the $a$ axis. One complete polarimetry scan for each detection configuration, i.e. 72 2D THz spectra, takes about 11 hours. The parallel detection shows weaker magnon upconversion signals than the perpendicular configuration, resulting in degraded quality of the polarimetry pattern due to interference with neighboring peaks. Fits to functions of the form of $\left|\cos^2\theta\sin\theta\right|$ and $\left|\cos^3\theta\right|$ are shown in \textbf{d} and \textbf{e} respectively.}
\end{figure}

\begin{figure}
	\centering
	\includegraphics[width=1\linewidth]{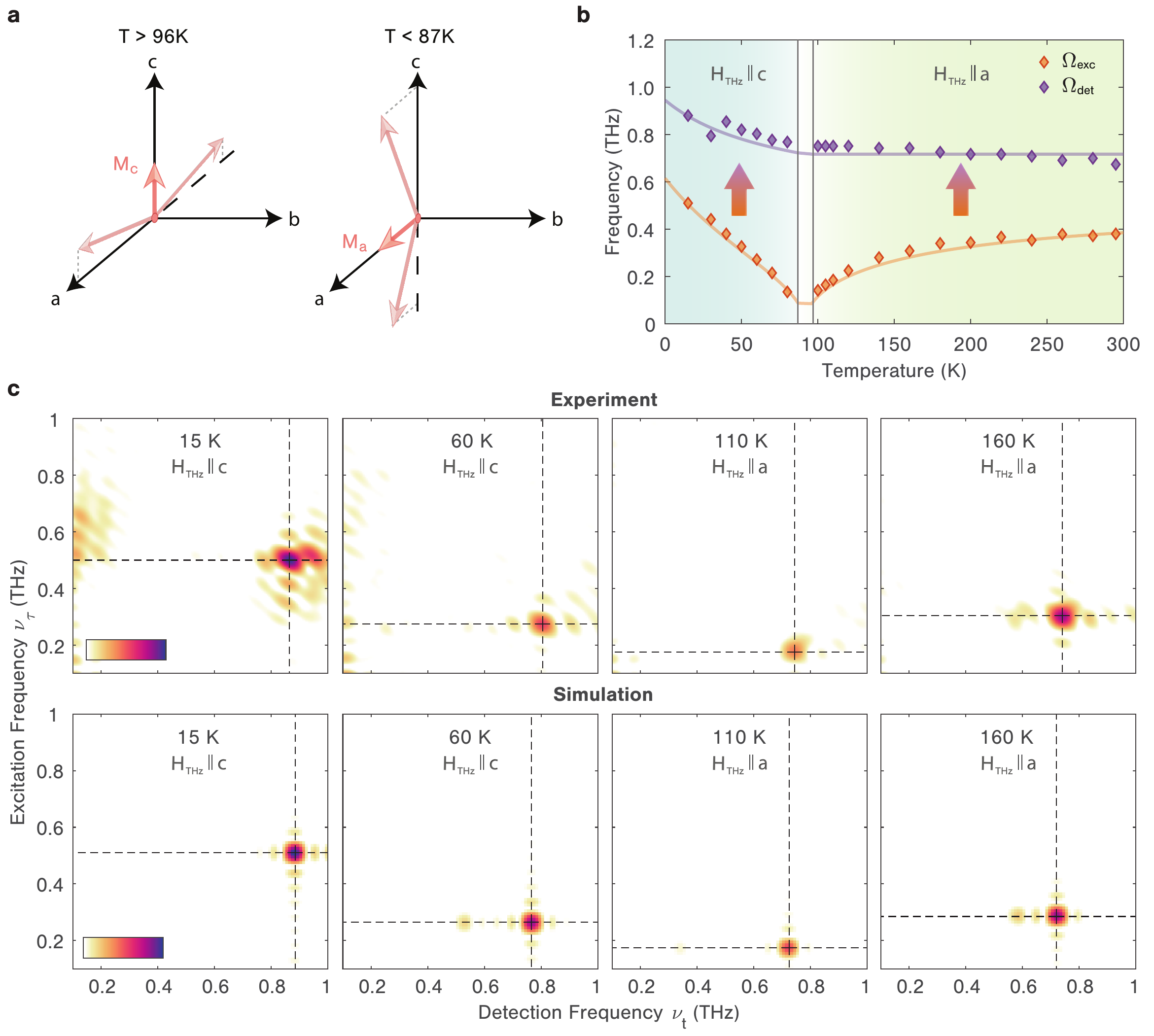}
	\caption{\label{fig:Fig3} 
	\textbf{Temperature dependence of magnon upconversion. a,} Schematic representation of the canted spin configurations for temperatures above and below the spin reorientation transition (87-96K) in EFO. \textbf{b,} Excitation (orange diamonds) and emission frequencies (purple diamonds) of magnon-magnon peak obtained from temperature-dependent perpendicular polarized 2D THz measurements are shown along with the frequencies of qAFM (purple solid) and qFM (orange solid) magnon modes derived from simulations of the spin dynamics using the LLG equation.  \textbf{c,} Raw experimental and simulated 2D THz spectra for selected temperatures above and below the SRT. Note that $\mathbf{H}_{\textup{THz}} \parallel a$ ($c$) above (below) the SRT drives the qFM mode. All spectra are in the perpendicular detection configuration ($\mathbf{H}_{\textup{det}} \perp $ $\mathbf{H}_{\textup{THz}}$).}

\end{figure}

\begin{figure}
	\centering
	\includegraphics[width=1\linewidth]{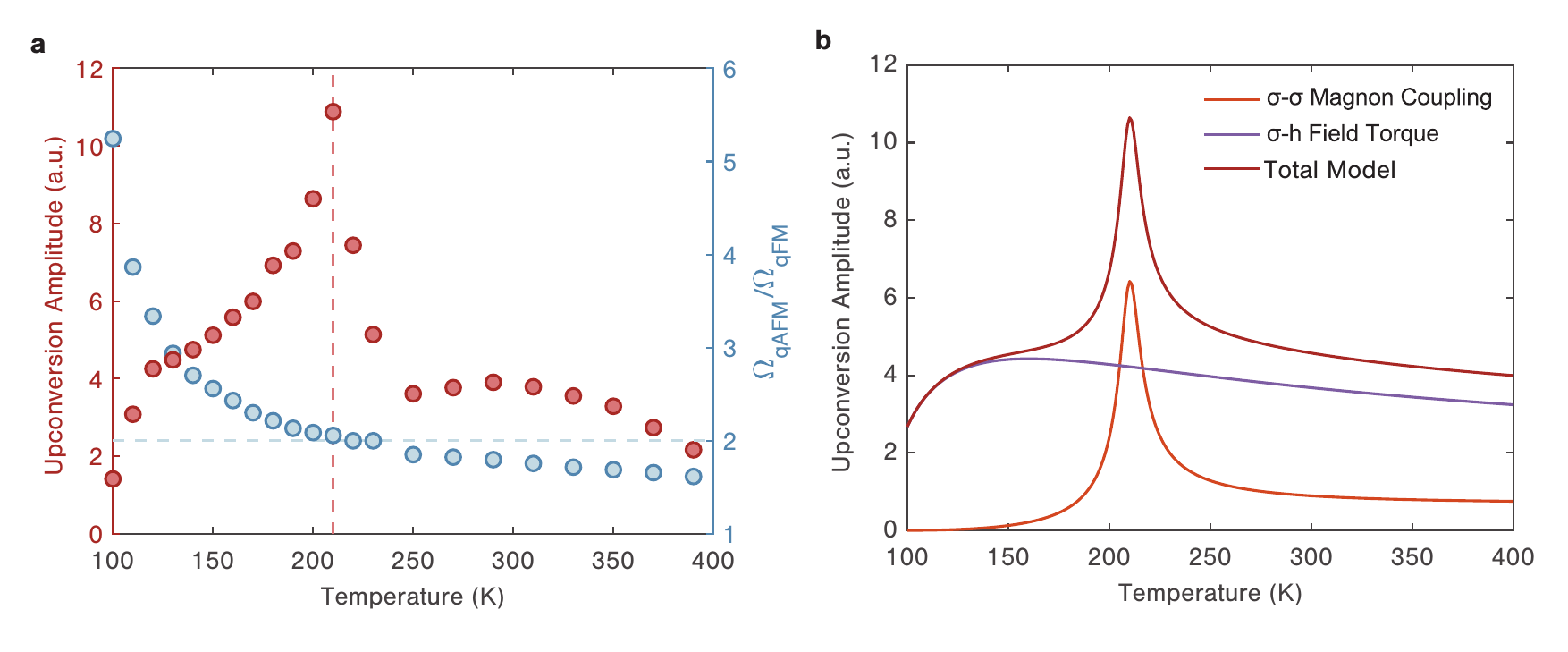}
	\caption{\label{fig:Fig4}  
	\textbf{Origin of the magnon upconversion process a,} Temperature dependence of the amplitudes of the upconversion signals (red dots) along with the ratios of the qAFM magnon frequency to the qFM magnon frequency above the SRT. The amplitude peaks at 210 K (red dashed line), where the qAFM mode frequency matches twice of the frequency of the qFM mode (i.e., $\Omega_{qAFM} = 2\Omega_{qFM}$) (blue dashed line). \textbf{b,} Estimated qAFM mode amplitudes driven by the magnon-magnon coupling (orange) and the magnon-field effect (purple line). The red line indicates the sum of the two contributions.}
\end{figure}

\FloatBarrier

\end{document}